\documentclass[9pt]{nature}
\usepackage{graphicx, caption, subcaption}
\usepackage{amssymb}
\usepackage{algorithm}
\usepackage{algpseudocode}
\usepackage{mathtools}
\usepackage{caption}
\newcommand{\supp}{Supplementary Information}

\usepackage{xcolor}

\title{People, Places, and Ties: Landscape of social places and their social network structures}

\author{Jaehyuk Park$^{1,2*}$, Bogdan State$^{3*}$, Monica Bhole$^{3}$, 
Michael C. Bailey$^{3\dagger}$, Yong-Yeol Ahn$^{1,4,5,6\dagger}$\\ 
\\
\normalsize{$^{1}$Luddy School of Informatics, Computing and Engineering, 
Indiana University Bloomington, IN 47408, USA}\\
\normalsize{$^{2}$Kellogg School of Management, 
Northwestern University, Evanston, IL 60208, USA }\\
\normalsize{$^{3}$Facebook, Menlo Park, CA 94025, USA }\\
\normalsize{$^{4}$Center for Complex Networks and Systems Research, 
Luddy School of Informatics, Computing and Engineering, 
Indiana University Bloomington, IN 47408, USA }\\
\normalsize{$^{5}$Network Science Institute, Indiana University, 
Bloomington (IUNI), IN 47408, USA}\\
\normalsize{$^{6}$Connection Science, Massachusetts Institute of Technology, 
Cambridge, MA 02139, USA }\\
\normalsize{$^*$ Contributed equally}\\
\normalsize{$^\dagger$Correspondence should be addressed to Michael C. Bailey (E-mail: mcbailey@fb.com)}\\
\normalsize{and Yong-Yeol Ahn (E-mail: yyahn@iu.edu)}\\ }

\begin{document}

\maketitle

\begin{abstract}
Due to their essential role as places for socialization, 
``third places\cite{oldenburg1999great}''---social places where people casually 
visit and communicate with friends and neighbors---have been studied by a 
wide range of fields including network science, sociology, geography, 
urban planning, and regional studies.
However, the lack of a large-scale census on third places kept researchers
from systematic investigations.
Here we provide a systematic nationwide investigation 
of third places and their social networks, by using Facebook pages. 
Our analysis reveals a large degree of geographic heterogeneity in the distribution 
of the types of third places, which is highly correlated with baseline demographics 
and county characteristics. Certain types of pages like ``Places of Worship'' 
demonstrate a large degree of clustering suggesting community preference or 
potential complementarities to concentration. 
We also found that the social networks of different types of social place 
differ in important ways: 
The social networks of `Restaurants' and `Indoor Recreation' pages 
are more likely to be tight-knit communities of pre-existing friendships whereas 
`Places of Worship' and `Community Amenities' page categories are more likely 
to bridge new friendship ties.
We believe that this study can serve as an important milestone for future studies 
on the systematic comparative study of social spaces and their social relationships. 
\end{abstract}

In January 2014, \textit{The New York Times} reported a conflict between a 
group of local senior citizens and management at a McDonald's in Queens, New York.  
The conflict centered around senior citizens who regularly spend 
time at the store socializing with one another 
without spending much on food beyond coffee and fries\cite{maslin2014fighting}.  
In other words, the McDonald's functioned as both a commercial establishment and
a community center, an example of what urban theorists
dub a ``third place''\cite{oldenburg1999great}, 
a social place where people can regularly visit and communicate with friends, 
neighbors, and even strangers, distintive from home (first place) 
and the workplace (second place).

Given their ubiquity and essential role as places for socialization, the idea 
of third places --- public spaces in particular --- has been studied by a 
wide range of fields including network science, sociology, geography, 
urban planning, and regional studies\cite{jacobs1961death, whyte1980spaces, 
lefebvre1992production, putnam2000bowling, Gieryn_2000, cho2011friendship, 
small2019role, dong2020segregated, hidalgo2020amenity}.  
Empirical studies show that public spaces such as parks and cafes attract 
people and initiate new social interactions\cite{whyte1980spaces, 
hickman2013third, hampton2015change}, 
and the accessibility of public spaces is associated with  
the level of social interaction and friendship formation between 
people around them\cite{abu1999housing, lund2003testing, small2009unanticipated}.
%Furthermore, such social places increase social capital of local 
%communities~\cite{oldenburg1999great, putnam2000bowling}. 
Furthermore, studies show that the rise and decline of local pubs in rural areas drive 
the corresponding change in the regional social capital and 
the level of social cohesion of the rural 
communities\cite{oldenburg1999great, putnam2000bowling, cabras2011industrial, 
cabras2017third}, especially during disasters or economic hardships\cite{browning2006neighborhood, 
  dynes2006social, airriess2008church, klinenberg2015heat}. 
On the other hand, the social infrastructure of an area also have a major 
impact on the residents' quality of life\cite{carley2001retailing, goodchild2008homes}. 
Hence, better understanding of social spaces and their function is critical to
improve the resilience of communities. 

Despite their importance in our social lives, 
the lack of systematically collected ``census'' of social places and their 
social networks has been a significant challenge for understanding
the nature and impact of third places. 
Although nationwide surveys (e.g, County Business Patterns by US Census)
produce information on local commercial establishments, 
they do not cover non-commercial (but essential) public spaces, 
such as community ammenities and outdoor parks.  
Most existing studies did not distinguish between various types of local places,
such as cafe, bars, and community ammenities,
impeding further comparative inquiries.
Finally, the interplay between the types of social places and 
the social networks which they facilitate has been mostly overlooked,
primarily due to the lack of appropriate data sources. 
For instance, what would be the differences in the social network framed around
an outdoor parks and another around a restaurant?
Would churches and pubs tend to facilitate similar types of social connections?

We address this challenge by investigating \emph{the landscape 
of social places and their friendship networks} geographically, and 
demographically, across the United States.
We use nationwide, de-identified, and aggregated data from Facebook Pages to measure the 
distribution of various third places, which allows us to 
present a systematic perspective of social spaces in the US. 
Furthermore, we use social network of Facebook followers of third place establishiments,
to examine the multi-faceted interaction between third places 
and social lives.

\section*{Results}

\subsection{Representativeness of Facebook pages} %{{{
\label{result:representativeness}
Facebook friendships provides a meaningful representation of people's social ties.
In the United States, the Facebook usage rate is not only high 
(69\% of adults), but also relatively constant across income groups, education groups, 
and racial groups among online US adults\cite{duggan2015social}.
Furthermore, previous studies of online social networks have found a significant association 
between self-reported friendships and Facebook 
friendships\cite{gilbert2009predicting, jones2013inferring, hampton2011social}.
However, while the representativeness of friendships in Facebook for 
real social relationships has been validated by previous studies and surveys,
the representativeness of Facebook Pages 
for offline third places across the US has not been thorougly examined.
To evaluate the representativeness of Facebook Pages,
we compare the number of Facebook social place pages 
to the number of third place establishments observed in 
the County Business Patterns (CBP) dataset (See Methods). 
Note that it covers only a fraction of social spaces 
and thus falls short for our systematic investigation.

\begin{figure*}[htpb] %{{{
    \footnotesize
    \centering
    \begin{subfigure}[t]{0.33\textwidth}
        \centering
        \includegraphics[width=\textwidth]{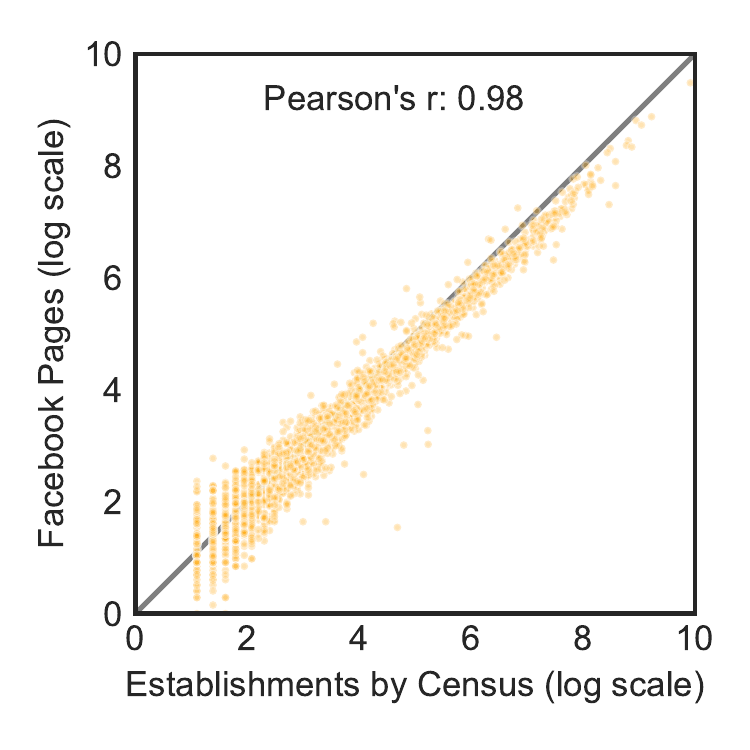}
        \caption{Restaurants}
        \label{fig:naics_restaurant}
    \end{subfigure}%
    \begin{subfigure}[t]{0.33\textwidth}
        \centering
        \includegraphics[width=\textwidth]{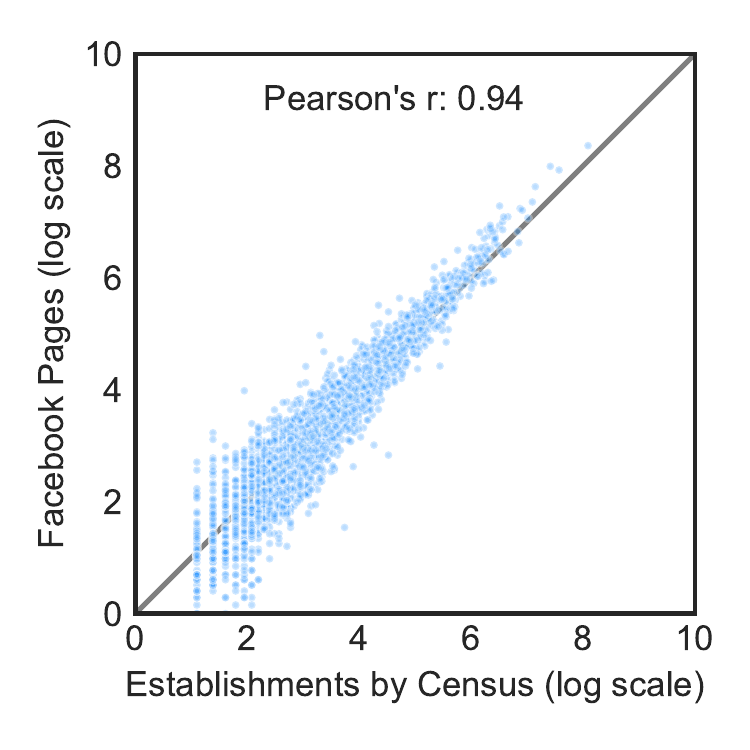}
        \caption{Places for Worship}
        \label{fig:naics_worship}
    \end{subfigure}
    \begin{subfigure}[t]{0.33\textwidth}
        \centering
        \includegraphics[width=\textwidth]{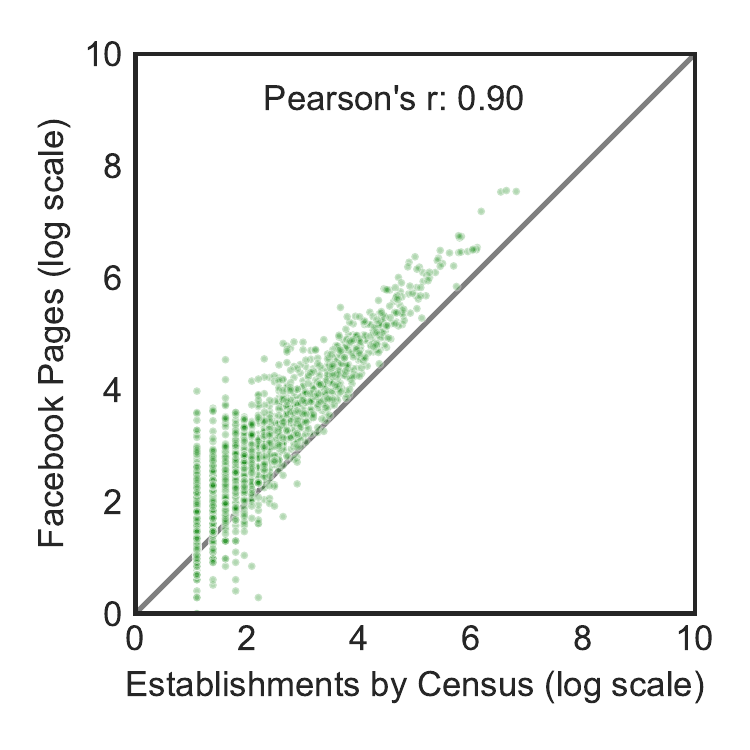}
        \caption{Bars and Pubs}
        \label{fig:naics_bar}
    \end{subfigure}%
 \caption{Correlation between the number of establishments 
   surveyed by the US Census and the number of Facebook's pages in similar categories
   (See \supp  for per-capita plots, which still show significant correlation). 
   All compared categories show
   significant correlations, which supports the representativeness of
   Facebook Pages to reveal the landscape of social places.}
\label{fig:fb_vs_naics}
\end{figure*}
%}}}

Figure~\ref{fig:fb_vs_naics}(a) to (c) show that there is a strong correlation 
between the logarithm of the number of social place pages and the estimated 
number of third place establishments per county in the CBP dataset. 
The Pearson's correlation coefficients range from 0.90 for Bars and Pubs to 0.98 for 
Restaurants.
We also check the correlations of the per-capita numbers (See \supp), 
which still show significant correlations across all categories.  
Despite there still exists potential discrepancy caused by overlapped 
or fake Facebook Pages, the strong correlations between the number of Facebook's pages and the 
government-estimated number of businesses 
is a promising sign that Facebook Pages cover a sizeable number of 
real-world businesses.
%}}}

\subsection{Geographic and Demographic Landscape of Social Places} %{{{
\label{result:landscape}
Facebook Pages provide a means for individuals to connect over shared interests in hobbies, 
causes, businesses, or celebrities, to mention but a few categories which form the 
subjects of Facebook pages. 
Facebook Pages includes physical locations, which are of local interests.
These pages on local places are arguably well-aligned with the concept of ``third places.'' 
Using an existing list of third place categories created by Jeffres 
et al.\cite{jeffres2009impact}, we systematically identify (See Methods) 
the pages of social places which connect people locally and may function as 
``third places.''

\begin{figure*}[htpb] %{{{
    \centering
    \includegraphics[width=0.8\textwidth]{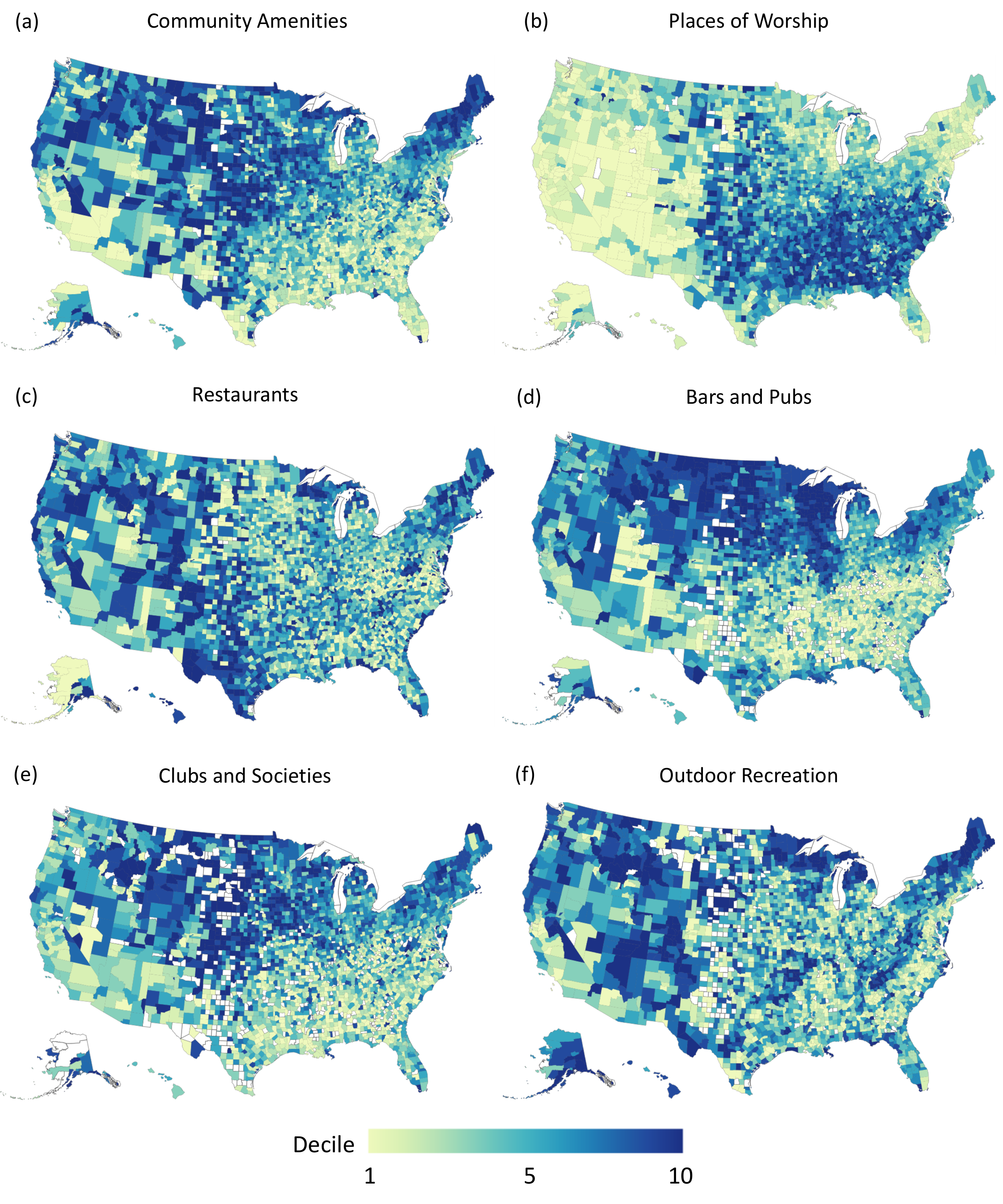}
    \caption{Prevalence maps of different social place categories 
      across the United States.      
      \textbf{(a)} Community Amenities, \textbf{(b)} Places of Worship,
      \textbf{(c)} Restaurants, \textbf{(d)} Bars and Pubs, \textbf{(e)} Clubs and Societies,
      and \textbf{(f)} Outdoor Recreation. 
      Prevalence is measured as the number of establishiments 
      per 1000 residents for each third place category. For visualization, 
      each US county is assigned into a decile based on its prevalence.
      There exist several third place types heavily concetrating on 
      a certain part of the US --- Places of Worship and Bars and Pubs.
      Also, the third places for Outdoor Recreation are distributed 
      around famous mountain areas, which shows the association between 
      natural environment and burgeoning types of social places. 
    }
    \label{fig:cat_dist}
\end{figure*}
%}}}

The maps of social place distribution exhibit that
there is significant geographic heterogeneity between third place 
categories.
`Bars and Pubs' and `Clubs and Societies' are more common in Midwest and Northeast, 
while Places for worship are more concentrated in the South.
On the other hand, Community Amenities and Restaurants are more common 
in the West and Northeast (Figure~\ref{fig:cat_dist}(a) and (c)). 
Finally, Outdoor Recreation are
distributed around famous mountain and lake areas, including the regions around 
Rocky Mountains, Appalachian, and Five Great Lakes (Figure~\ref{fig:cat_dist}(f)).

\begin{figure*}[htpb] %{{{
    \centering
    \includegraphics[width=\textwidth]{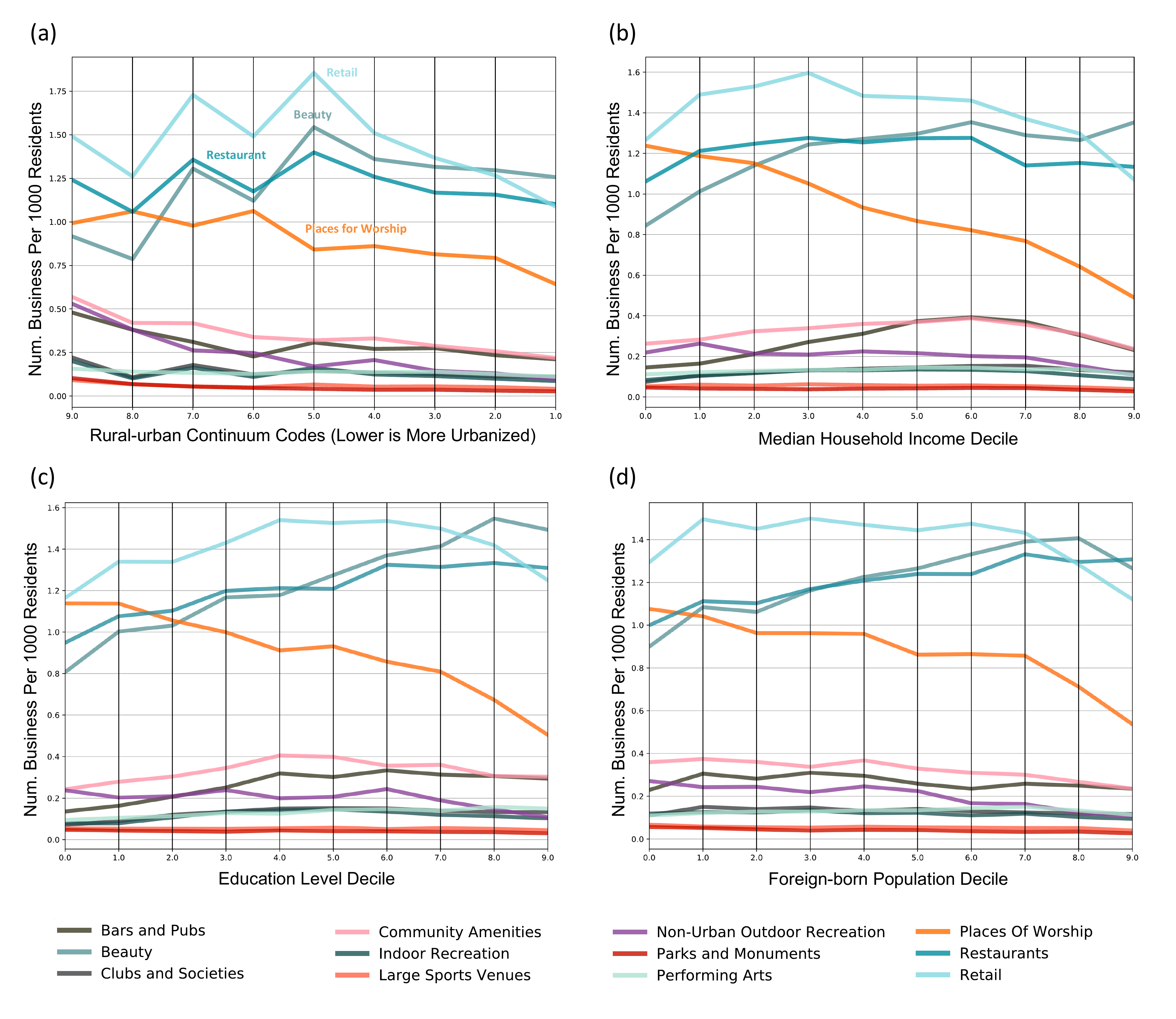}
    \caption{Comparison of the median numbers of each social places per 1000 residents 
      in US counties across different levels of \textbf{(a)} urbanization, 
      \textbf{(b)} median household income, \textbf{(c)} education, 
      and \textbf{(d)} foreign-born population.
      Many third place types show distinct pattern based on 
      demographic characteristics, such as Places of Worship having a dramatic
      decreasee with increasing urbanization, income, education, and foreign-born
      population.
    }
    \label{fig:cat_parallel}
\end{figure*}
%}}}
Community characteristics and demographics play a large role in which 
establishments are created and persist. 
To better understand the relationship between demographics and prevalence of 
third places, we look at four characteristics and 
how they influence third place prevalence: 
urbanization, income, education, and foreign-born population.

The results, shown in Figure~\ref{fig:cat_parallel}, 
reveals that prevalence of the four largest categories --- 
Retail, Beauty, Restaurants, and Places of Worship --- 
indeed varies with regional and demographic characteristics. 
Counties in the middle of the RUCC urbanization scale have the highest prevalence 
of retail stores, beauty shops, and restaurants (Figure~\ref{fig:cat_parallel}(a)). 
Places of Worship, Community Amenities, and Outdoor Recreations, 
in particular, are less prevalent in more urban counties.

The prevalence of Places of Worship shows a stark decrease with increasing income levels.
The counties in the lowest income decile tend to have about 2.5 times more places 
of worship per capita than the counterparts in the highest decile.  
By contrast, the numbers of beauty places in rich counties are higher than 
the number in poor counties in general.  
Other categories, such as Community Amenities, Bars and Pubs, Retail, 
and Restaurants, 
show the inverted-U shape, where the middle-class counties have more of 
those social places than either richer and poorer counterparts.  

As expected from the correlation between income and other demographic characteristics,
the prevalence of Places of Worship shows a similar decreasing pattern 
with education level and foreign-born population.
Community Amenities are more prevalent in 
the regions having medium education level and lower portion of 
foreign-born population, while the prevalence of Beauty and Restaurants
are correlated with both the level of education and the portion of 
foreign-born population. 
The heterogeneous prevalence across the types of
social place, depending on demographic characteristics of region, 
reveals the importance of comparative studies, in the future, to 
understand the role and effect of social places properly.

%}}}

\subsection{Social Networks of Social Places}
\label{result:social_network}

If types of available social places are strongly associated with
geogrzphic and demographic characteristics,
how about the social network around each type of social place?
%Given long-established theories about differences in the topology of 
%social networks across the United States (e.g.~\cite{putnam2000bowling}), 
Are there certain characteristics of social networks that are
associated with each type of social place?
For instance, how does the social networks of people who frequent 
a Restaurant differ from those who frequent a Place of Worship?
We answer this question by leveraging the Facebook friendship network 
of users who follow of a particular page in Facebook Pages, which we call
``follower friendship network''. 
This approach creates a user-to-user graph for each page,
which allows us to analyze the structure of social relationships embedded 
in each social place. 

We first characterize the social network topologies 
around all third place categories by measuring various network features. 
We sampled 2,500 pages having between 50 and 50,000 followers
--- to exclude extreme pages --- at random, from 
each of the twelve third place categories.
Then, we compute multiple topological statistics of 
the social network of each sampled page
(See Methods for more information on the extraction process). 
In particular, following a previous study to characterize 
social networks of US colleges\cite{overgoor2020structure}, 
we compute the following 18 network measurements 
with respect to the user-to-user friendship graph connecting followers of each page: 
density, number of edges, number of nodes, average degree, 
average clustering coefficient\cite{watts1998collective}, 
average degree assortativity\cite{newman2003mixing}, 
degree variance, average path length within the largest connected component, 
algebraic connectivity\cite{fiedler1973algebraic}, 
modularity of modularity-maximizing partition\cite{clauset2004finding}, 
and number of $k$-Cores\cite{bollobas2001cambridge} and 
$k$-Brace~\cite{ugander2012structural} for $k \in \{2, 4, 8, 16\}$.
These measures examining various aspects in network structure 
allow us to capture the fragmentation and diversity of social networks, 
which we expect to differ by type of social place.

\begin{figure*}[htpb]
 \includegraphics[width=\textwidth]{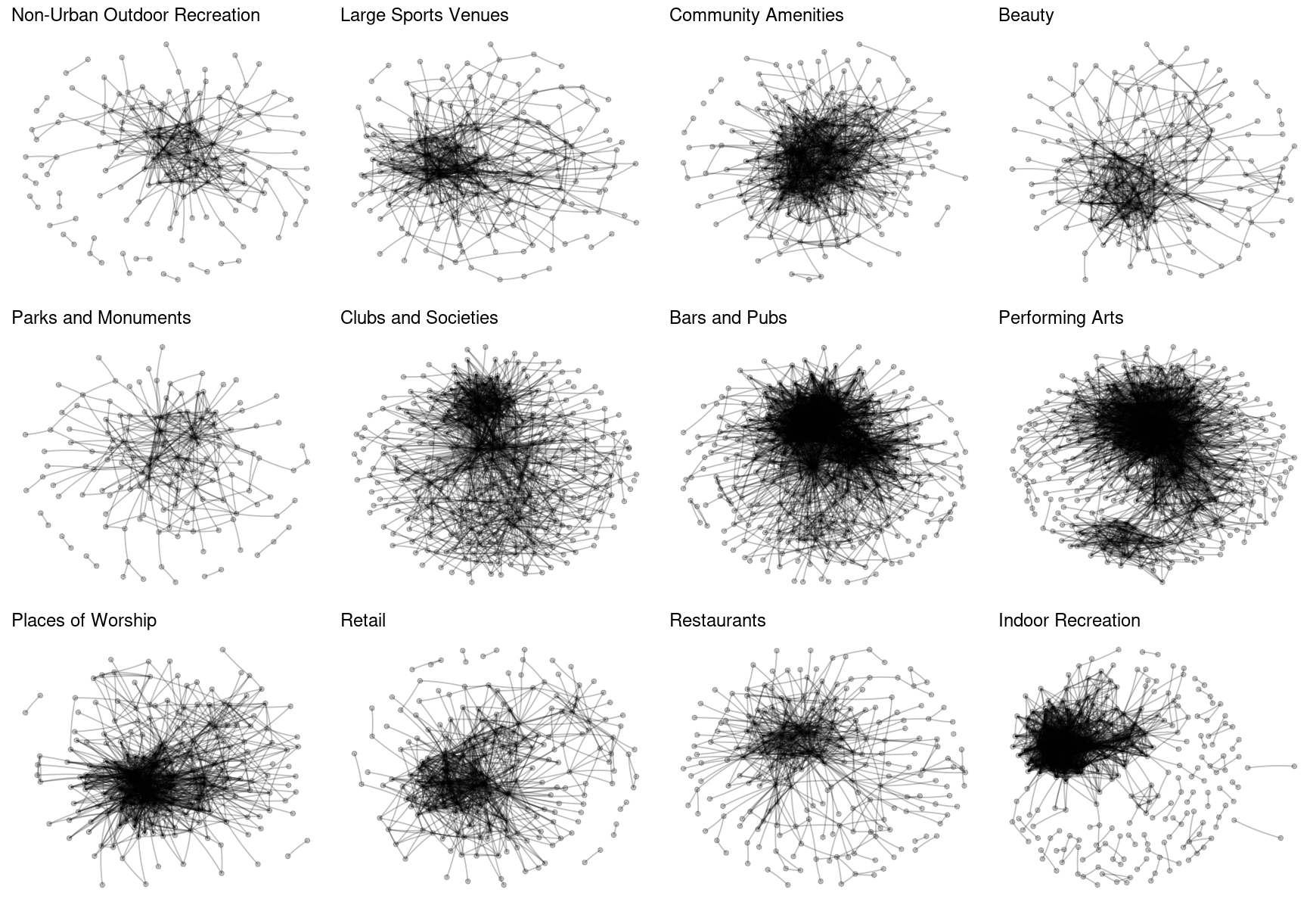}
 \caption{Typical networks for each third place category. 
    Plot shows page follower networks that are nearest to the per-category mean 
    ranks. Distance defined as the L2-norm against the per-category mean rank 
    across all features, weighted by feature importance.
    The structure of frienship networks can provide hints about the functional role of 
    the third places in our social lives. For instance, the prevalence of
    dyads and triads separated from a giant component (e.g, Restaurants)
    coincide with our intuition that restaurant visitors
    are likely to come with a small group of
    their existing friends, while less likely to know other people in the place.
    Also, the existance of core-periphery structure strongly suggests 
    the existence of ``regulars''. 
  }
 \label{fig:representative}
\end{figure*}

We first extract the `most representative' network for each third place 
category and visualize them (See Figure \ref{fig:representative}). 
%The distance is the L2-norm from this per-category average, 
%weighted by feature importance (as in Figure~\ref{fig:feature-importance}). 
We find that social networks for Parks and Monuments and Outdoor Recreation are 
sparser, less-centralized, while those for Clubs and Societies, 
Bars and Pubs, and Performing Arts venues are highly connected,
having multiple cores. 

The structures of the representative friendship network reveals
about the functional role of the third place type in our social lives.
The coverage of the largest connected group of each friendship network may imply
whether the type of third place is more for small groups of existing friends or 
for constructing a bigger community by bridging gaps between strangers.
The representative networks that exibit many independent dyads and triads---those 
of Outdoor Recreation, Indoor Recreation, Restaurant, Parks and 
Monuments, for instance---potentially indicate that these social places are where people visit 
with a small group their existing friends.
On the contrary, the social networks having a large connected subgraph covering most members
may suggest that the social places where many visitors are likely to know each other 
or to be introduced to each other. 
Also, the existence of clear division between a group of densely-connected 
core members and the loosely-connected other members, called core-periphery 
structure, hints at the presence 
of ``regulars''; The friendship networks of 
the places having a group of core members who are densely connected to each 
other --- such as Places of Worship, Bars and Pubs, Community Amenities, and 
Indoor Recreation --- is consistent with the story that there are regulars who 
frequently visit the place, while other non-regulars are likely to be a friend 
of one of the regulars. 

\begin{figure*}[htpb] %{{{
    \footnotesize
    \centering
    \includegraphics[width=0.8\textwidth]{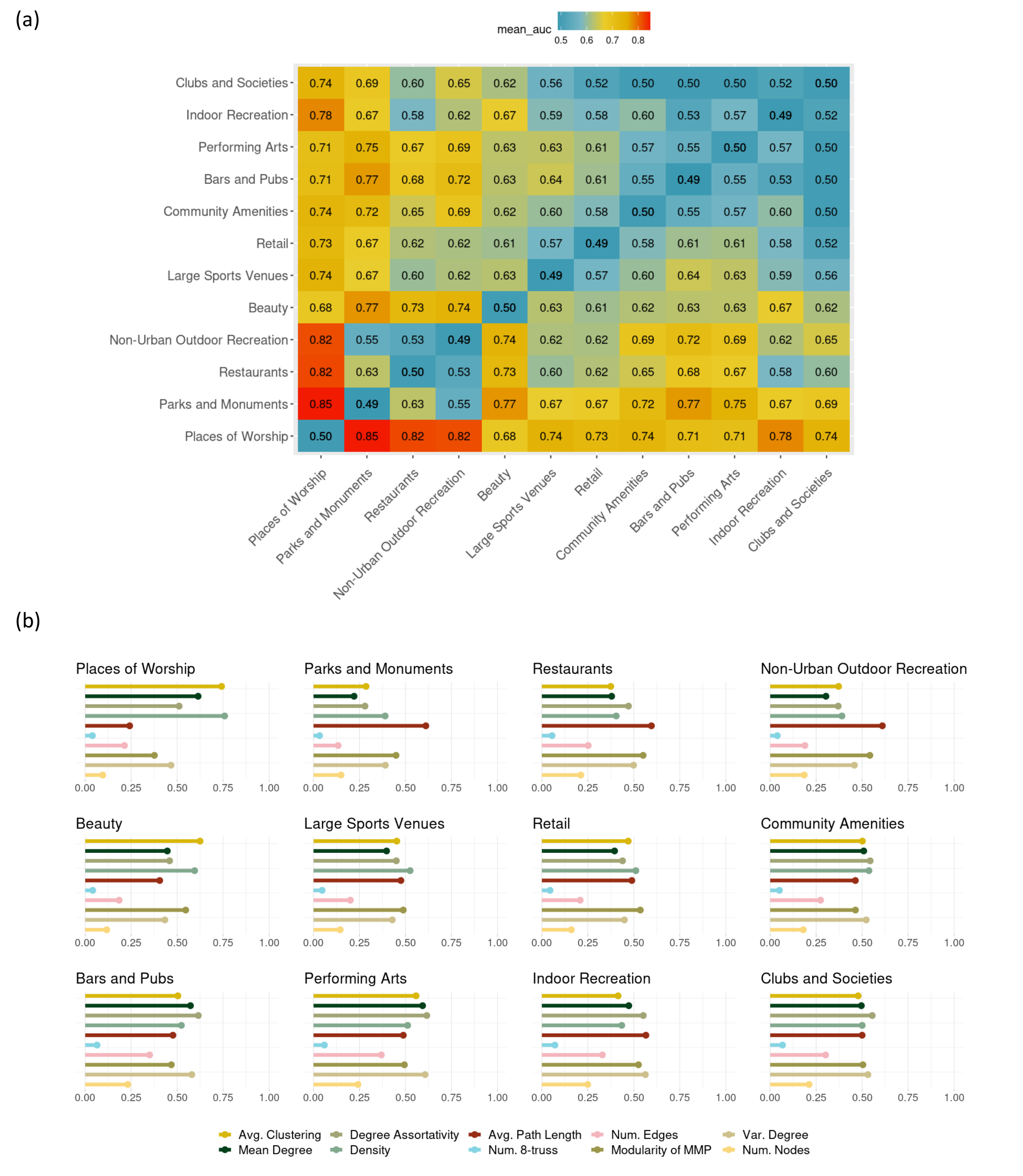}
    \caption{\textbf{(a)} Similarity (AUC) matrix for third place categories
    based on their social network structure. ROC--AUC of 0.5 represents the
    maximal similarity (indistinguishable) and ROC--AUC of 1.0 represents
    the minimal similarity (perfectly distinguishable). 
    The matrix shows roughly two groups: one with Parks and Monuments, Resturants,
    and Outdoor Recreation, and the other with most other categories, 
    while Places of Worship exhibit the most distinct structure.
    \textbf{(b)} Descriptive statistics of the structure of 
    friendship network associated with different third place categories.
    The statistics are the top ten features by importance, which is presented in \supp. }
\label{fig:roc-auc}
\end{figure*}

We then quantify concretely the structural difference between 
third place categories by measuring network dissimilarity 
between each pair of third place categories. 
Here, we measure similarity between two categories as 
\emph{the difficulty of classification}. 
If a classifier cannot separate two groups of samples easily,
we consider that the graphs that produced them are similar. 
This approach of measuring similarity with a prediction task has been 
used in recent studies\cite{gentzkow2019measuring, bertrand2018coming,
overgoor2020structure}.
For each pair of third place categories, we train a random forest classifier.
We train the sampled social networks of the two categories 
using the 18 aforementioned topological characteristics as the features
(See Methods for detail information). 
As a result of all possible pairs of the twelve categories, 
we obtain the cross-validated area under the curve (AUC) of the model for 
each pair of third place categories, as a measure of similarity distance 
between the cateogories.
In other words, if the AUC is close to 0.5, the two 
categories cannot be easily distinguished, which can be considered as similar. 
On the other hand, if the AUC is close to 1.0, the two categories are 
easy to distinguish, which can be considered as different.

The ROC--AUC distance matrix, as shown in Figure \ref{fig:roc-auc}(a), reveals that 
it is generally possible to distinguish between follower networks corresponding to 
different third place categories. Also, the matrix shows two clusters of third place
categories having similar network structures to each other; 5-category 
cluster (Clubs and Societies, Indoor Recreation, 
Performing Arts, Bars and Pubs, and Community Amenities), and 
3-category cluster (Outdoor Recreation, Resturants, Parks and Monuments).
Places of Worship display the most distinctive pattern, indicating
that social network around the Places of Worship is highly 
distinguishable from most other categories. 
Figure \ref{fig:roc-auc}(b) reveals the most informative characteristics of
the friendship networks in distinguishing them across categories. 
We check the descriptive statistics for the top 10 features
on average for each type of third place, 
ranked in order of importance across all classifiers (For the list of feature
importance, See \supp). 
Aligning with the distant matrix, the topological features show clear 
similarity between the types within each group, while difference between 
the groups. Parks and Monuments, Resturants, and Outdoor Recreation 
are distinctive based on their Avg. Path Length, 
while not so regarding clustering measurements, 
such as Avg. Clustering, Degree Assortativity, and Density.
At the same time, the third places in the 5-category cluster is distinguished 
from the others --- Beauty, Large Sports Venues, and Retail ---
in terms of Variation of Degree, which is probably related to 
their core-periphery structure.

It is noteworthy that the similarity of social network between third place
categories is not much correlated with the types of activity or behavior 
in the places.
For instance, social activities and behaviors in Resturants 
are more similar to those in Bars and Pubs than Community Amenities. 
However, in terms of the topology of friendship network, 
Bars and Pubs is closer to Community Amenities
than Resturants, which is aligned with previous studies about the social role
of pubs in rural area\cite{cabras2011industrial, hickman2013third, cabras2017third}. 
Hence, our results here imply that it is essential to take the cultural and 
environmental factors into account, beyond the type of activity,
when studying the third places.

\section*{Conclusion}

In this study, we present a systematic measurement on the prevalence 
of various third places across the United States, by leveraging Facebook Pages dataset. 
Our results reveal differences in geographical distribution of social places 
(e.g. places of worship in the South, and bars in the Midwest), 
as well as the distribution with respect to demographic features,
including the levels of urbanization, income, education, and foreign-born population. 
Our results also reveal that different kinds of third places draw upon (or facilitate)
heterogeneous social network structures among the ``followers'' of their pages. 
For instance, Places of Worship tend to be associated with 
social networks that are highly clustered and feature short path lengths. 
Parks and Monuments, by contrast, have low clustering and 
longer path length, indicating a more sparsely-organized networks.
Certain third place categories 
(e.g. Indoor Recreation and Clubs and Societies) 
have more similar social networks than others, 
and social networks systematically differ across categories. 

We expect that our study marks an important milestone towards 
the understanding of our social infrastructure and their roles 
in our society by exploring a unique dataset that 
covers all major types of social spaces and spans a whole country. 
Since our findings are decidedly descriptive, 
more work is required to ascertain the extent to which certain kinds of 
third places also help create and maintain social ties, as well as 
the extent to which third places benefit from existing social networks. 
A dynamic view is thus called for in future research, examining the interplay 
between social ties and third places.

There are several limitations of this study.
First, even with the strong correlation between 
the prevalence for third place categories in Facebook Pages and 
county-level CBP statistics, we cannot completely rule out 
the potential existence of systematic biases in our dataset. 
For instance, the delay between the actual opening (closing) of
establishments and the creation (deletion) of their Facebook Pages
may affect our findings. Second, given the existence of a digital divide, 
our observations may have been affected by the preferred social places 
for more online-friendly generations.
For example, our ranking of the total number of Facebook pages 
(See \supp) are more consistent with the rankings 
presented in a previous survey based on three college town areas 
in Massachusetts\cite{mehta2010third}, 
than the social place ranking in another study based on a national telephone 
survey\cite{jeffres2009impact}. 
Even if our study could pinpoint the \emph{existence} of social places, 
it does not necessarily capture their \emph{size} or \emph{usage}. 
The number of places of worship may not reflect 
the number of people attending services, since the size of congregations can 
vary a great deal. Leveraging other datasets may address this issue in the future. 

During the last couple of decades, we have experienced a dramatic change 
in our social lives due to the emergence of a new type of social spaces 
--- social media. As we become more familiar with online social places 
such as Facebook or Twitter, this new social space becomes more related 
and embedded into the existing offline social places.  
As our favorite social places create their own websites and Facebook
Pages to interact with customers online and offline social places
become more interdependent. 
Hence, studying online and offline interactions can help us to understand
the similarities and differences between these different spaces. 
Furthermore, examining the changes in the landscape of the social spaces --- 
particularly with respect to the COVID-19 pandemic and increasing 
interconnectedness of offline and online social lives --- 
as well as studying the associations between the abundance of 
social places and the characteristics and dynamics of local 
communities will be fruitful future studies. 

\begin{methods}
\label{sec:methods}
\subsection{Matching CBP cateogories with the corresponding categories in Facebook Pages}
The County Business Patterns (CBP) dataset is created by surveying firms around 
the US and categorizing them into place categories by self-assessments to estimate 
the number of establishments in each category.

We manually matched a subset of social place categories with their corresponding 
North American Industry Classification System (NAICS) codes. 
For each matched categories, we compare the estimated number of establishments 
in the CBP dataset to the number of Facebook pages in the category. 
The matching table between our social place categories and NAICS codes is 
presented in \supp.

\subsection{Extraction of third place pages in Facebook Pages}
Since Facebook pages can fall into a number of categories, we create a data-driven 
taxonomy of Facebook Pages that represents social places.
In doing so we made use of a dataset of over 6 million local pages with between 50 and 50,000 
US followers, people who clicked ``like'' or ``follow'' on the page to follow its posts and updates. 
In Facebook Pages, page administrators choose up to 3 page categories for their pages 
--- for instance, a page may be identified as both ``AMERICAN\_RESTAURANT'', ``RESTAURANT'', 
and ``FOOD'' in the Facebook page category. 
Using a set of page categories for each page, we trained a 
word2vec model\cite{mikolov2013distributed}\footnote{In particular, we applied the implemented 
 version of word2vec model in the \texttt{gensim} package\cite{rehurek_lrec}}. 
For every broad category indicated by a previous work\cite{jeffres2009impact}, 
we choose a Facebook Pages' category that could reasonably represent 
the broader category. For instance, ``AMERICAN\_RESTAURANT'' 
was chosen for the ``restaurants'' categories.

The top 300 terms, in terms of their distance in the embedded space were then examined for each 
category, with the research team filtering categories that were judged to not fit 
the notion of third place. 
For instance, categories such as ``COMPETITION'' were excluded for not indicating a place, 
whereas ``MEDICAL\_HEALTH'' was excluded due to the ambiguous nature of the category, and so on. 
We also removed ``SCHOOLS'' from our consideration as they represent workplace 
(``second places'', rather than ``third places'') for students. 
Inspection of the data further led us to combine ``Restaurants'' and ``Cafes'' into a 
single category, as we did with Community Centers, Senior Centers, and Libraries. 
Finally, we opted for a different categorization of recreation venues --- 
instead of distinguishing only between community centers, indoors recreation, and outdoors 
recreation, we focused on Community Amenities, Performing Arts venues, Parks and Monuments, 
Indoor Recreation, Non-Urban Outdoor Recreation, and Large Sports Venues.  

\begin{table}
 \scriptsize
 \centering
 \begin{tabular}{|c|l|}
 \hline
 \textbf{Third Place Types} & \textbf{Facebook Pages Category} \\
 \hline
 Places of workship & 
 \textbf{BUDDHIST\_TEMPLE, CHURCH, HINDU\_TEMPLE, CATHOLIC\_CHURCH, MOSQUE,} \\
  & \textbf{SYNAGOGUE}, CHRISTIAN\_SCIENCE\_CHURCH, SEVENTH\_DAY\_ADVENTIST\_CHURCH, \\
  & ASSEMBLY\_OF\_GOD, CHURCH\_OF\_CHRIST \\ 
   \hline
 Restaurants & 
 \textbf{AMERICAN\_RESTAURANT}, ITALIAN\_RESTAURANT, EASTERN\_EUROPEAN\_RESTAURANT, \\ 
  & JAPANESE\_RESTAURANT, MONGOLIAN\_RESTAURANT, BUNSIK\_RESTAURANT \\ 
   \hline
 Bars & 
 \textbf{TOPIC\_BAR}, WHISKY\_BAR, TIKI\_BAR, IRISH\_PUB, BEER\_BAR, WINE\_BAR \\ 
   \hline
 Community Amenities &
 \textbf{COMMUNITY\_CENTER, MODERN\_ART\_MUSEUM, SENIOR\_CENTER, TOPIC\_LIBRARY,}\\
 & \textbf{TOPIC\_MUSEUM}, COMMUNITY\_MUSEUM, HISTORY\_MUSEUM, PUBLIC\_GARDEN, \\
 & COMMUNITY\_GARDEN, ART\_MUSEUM \\ 
   \hline
 Performing Arts &
 \textbf{LIVE\_MUSIC\_VENUE, SYMPHONY, THEATRE, TOPIC\_CONCERT\_VENUE}, COMEDY\_CLUB,\\
 & OPERA\_HOUSE, PERFORMANCE\_ART, AUDITORIUM, PERFORMING\_ARTS \\ 
   \hline
 Parks and Monuments &
 \textbf{AQUARIUM, ARBORETUM, DOG\_PARK, MONUMENT, PICNIC\_GROUND, PROMENADE, ZOO}, \\
 & WATER\_PARK, PUBLIC\_GARDEN, STATUE\_FOUNTAIN, RESERVOIR, WILDLIFE\_SANCTUARY \\ 
   \hline
 Indoor Recreation &
 \textbf{BOWLING\_ALLEY, MOVIE\_THEATRE}, DRIVING\_RANGE, DRIVEIN\_MOVIE\_THEATER, \\ 
   & HAUNTED\_HOUSES, KARAOKE, GO\_KARTING \\ 
   \hline
 Non-Urban Outdoor Recreation &
 \textbf{HIKING\_TRAIL}, TOPIC\_MOUNTAIN, ATV\_RENTAL\_SHOP, FISHING\_STORE, \\
 & GLACIER, FISHING\_CHARTER \\ 
   \hline
 Large Sports Venues &
 \textbf{FOOTBALL\_STADIUM}, FIELD, BASEBALL\_STADIUM, RUGBY\_PITCH, SOCCER\_STADIUM, \\
 & CRICKET\_GROUND \\ 
   \hline
 Clubs and Societies &
 \textbf{GYM, SOCIAL\_CLUB}, BASKETBALL\_COURT, BOWLING\_ALLEY, FENCING\_CLUB, \\
 & SALSA\_CLUB, BOXING\_STUDIO \\ 
   \hline
 Retail & 
 \textbf{SHOPPING\_MALL}, CLOTHING\_STORE, SPORTSWEAR\_STORE, POPUP\_SHOP, \\
 & FROZEN\_YOGURT\_SHOP, FASHION\_DESIGNER \\ 
   \hline
 Beauty & 
 \textbf{BARBER\_SHOP, HAIR\_SALON}, TANNING\_SALON\_SUPPLIER, COSMETICS\_BEAUTY\_SUPPLY, \\ 
   & SKIN\_CARE\_SERVICES, HAIR\_REPLACEMENT, MASSAGE \\ 
   \hline
 \end{tabular}
   \caption{Third place categories and matched Facebook Pages categories. 
   Top categories for each type are presented in bolded text. 
   See \supp for full listing.}\label{tab:cat_ex}
 \end{table}

The full results of our categorization exercise are shown in \supp, 
which shows the list of the twelve social place 
categories and the examples of the Facebook page categories corresponding 
to each social place category. 
As a result of the previously-outlined procedure, 453 Facebook page categories
matched with one of the 12 social place ``super-categories.'' Altogether
these ``social place pages'' account for 37.3\% of all US local Facebook pages 
with between 50 and 50,000 fans. % (Table~\ref{tab:cat_list}). 

Based on this table, we then filtered and matched the local places in Facebook Pages 
belonging each of the following twelve social place categories: places of worship,
restaurants, bars, community amenities, performing arts, parks and monuments,
indoor recreation, non-urban outdoor recreation, large sports venues, 
clubs and societies, retail, and beauty. 
We use the number of pages in a category per one thousand residents, 
based on the county-level population estimates 
in 2018 by US Census, which allows us to control for the population of US counties. 
Since each page in Facebook pages has up to three categories, we adjust the weight
of a page for counting, by dividing one by the number of categories that a page has.
For instance, if one page has two categories, 0.5 pages are counted for the two 
categories, respectively.

\subsection{Demographic records of counties}
As a proxy of the urbanization level of a county, 
we use The 2013 Rural-Urban Continuum Codes (RUCC) created by the 
Office of Management and Budget (OMB) of the United States. 
RUCC is a classification scheme to distinguish metropolitan counties by the population 
size of their metro area, and nonmetropolitan counties by degree of 
urbanization and adjacency to a metro area.  
Under this scheme, US counties are coded from 1 
(Counties in metro areas of 1 million population or more) 
to 9 (Completely rural or less than 2,500 urban population, not adjacent to a metro area), 
based on the population and adjacency to a metro area.  
For regional income, we utilized median household income records for each county 
from 2018 American Community Survey (ACS).
Also, for education and foreign-born population, we use
the number of people aged 25 or older who have higher than college degree in each county
and the proportion of foreign-born residents in each county, respectively,
from 2006-2010 ACS. 
To visualize demographic patterns in third place distributions,
we compute the number of third places per 1000 residents for each county. 
We then compute the median for each RUCC code and for each decile in terms of 
the income, education, and foreign-born population proxies.

\subsection{Extracting social graphs of Facebook Pages}
For each of 2,500 randomly sampled Facebook Pages of 
each of the twelve third place categories, 
we extract the friendship network of the users 
who are the fan of the page. 
Then, we map the page to its corresponding third place category, 
and calculate the topological features of our interests,
associated with the category.
Finally, after mapping the topological features into each category,
all network information is discarded. 
Since this study is only interested in the topological structure 
of each third place category, 
the entire process constitutes appropriate aggregations for each step,
which carry no personally identifiable information.

\subsection{Measuring similarity between social networks}
We trained 66 binary random forest classifiers --- 
one for each possible pair of different third place categories. 
The classifiers were trained under 10-fold cross-validation, 
using the \texttt{scikit-learn} Python package\cite{pedregosa2011scikit} 
to distinguish between page follower networks coming from either one of the paired categories. 
Feature importance, averaged across all classifier runs, 
reveals average clustering, mean degree, 
and degree assortativity as the most discriminative features. 
The list of importance for all 18 features are presented in \supp. 
This ordering suggests that the classifier is picking up on non-trivial structural 
differences in the make-up of page follower networks, rather than simply 
focusing on the number of followers (10/18 in terms of average importance) or density (4/18). 

\end{methods}

\bibliography{main}

\end{document}